\documentclass[10pt,aps,preprint,pra,twocolumn]{revtex4}
\usepackage{amsmath}
\usepackage{graphicx}
\usepackage{amssymb}
\usepackage{multirow}

\begin{document}

\title{Electronic and optical properties of bilayer blue phosphorus}
\author{Y. Mogulkoc}
\affiliation{Department of Engineering Physics, Faculty of Engineering, Ankara University, 06100, Tandogan, Ankara, Turkey}
    
\author{M. Modarresi}
\affiliation{Department of Physics, Ferdowsi University of Mashhad, Mashhad, Iran}
\affiliation{Department of Physics, Izmir Institute of Technology IZTECH, Izmir, Turkey}

\author{A. Mogulkoc}
\email{mogulkoc@science.ankara.edu.tr}
\affiliation{Department of Physics, Faculty of Sciences, Ankara University, 06100, Tandogan, Ankara, Turkey}
\author{Y.O. Ciftci}
\affiliation{Department of Physics, Faculty of Sciences, Gazi University, 06500, Teknikokullar, Ankara, Turkey}
\date{\today}

\begin{abstract}
We investigate the electronic and optical properties of monolayer and stacking dependent bilayer blue phosphorus in the framework of density functional theory (DFT) and tight-binding approximations. We extract the hopping parameters of TB Hamiltonian for monolayer and bilayer blue phosphorus by using the DFT results. The variation of energy band gap with applied external electric field for two different stacks of bilayer blue phosphorus are also shown. We examine the linear response of the systems due to the external electromagnetic radiation in terms of the dielectric functions in the DFT theory. The relatively large electronic band gap and possibility of exfoliation form bulk structure due to weak interlayer coupling, make blue phosphorus an appropriate candidate for future electronic devices. 
\end{abstract}

\maketitle
\section{Introduction}
After the discovery of graphene \cite{novoselov2004electric}, other 2D nano-structures were predicted theoretically \cite{xu2013large,modarresi2015effect,kaloni2015electrically} and synthesized in laboratory \cite{tao2015silicene,davila2016few,zhu2015epitaxial}. Among these, the monolayer black phosphorus, 2D puckered structure of phosphorus, which was also successfully fabricated in laboratory \cite{reich2014phosphorene,liu2014phosphorene} and studied with several theoretical works \cite{rudenko2014quasiparticle,rudenko2015toward,rodin2014strain,pereira2015landau,zhou2015landau,ezawa2014topological,PhysRevB.93.085417}. Moreover, another 2D structure of phosphorus with A7 phase which is known as blue phosphorus, is confirmed to be as stable as 2D black phosphorus due to the absence of imaginary frequencies in phonon spectrum \cite{zhu2014semiconducting,PhysRevLett.113.046804,aierken2015thermal}. In the meanwhile, there are few number of theoretical studies on buckled structure of phosphorus \cite{aierken2015thermal,ding2015structural,ghosh2015electric}. Also it was shown the blue phosphorus is stable under substitution of light non-magnetic atoms \cite{sun2015first}. Recently the quantum spin-Hall states have been predicted in bilayer black phosphorus \cite{zhang2015stacked}. 
The carbon atoms in the graphene have $sp^{2}$ hybridization which leads to in-plane $\sigma$ and an out-of-plane $\pi$ states in graphene plane. In the case of phosphorus the hybridization is $sp^{3}$ which is caused by the extra valence electron. The $sp^{3}$ hybridization leads to the out-of-plane atomic position and the buckled structure in phosphorus 2D nano structures \cite{PhysRevB.92.104104}. The electronic band in graphene are mostly arised from the atomic $p_{z}$ orbital in the $\pi$ states perpendicular to the graphene plane. As a result, the simple single tight-binding model works for low energy states around the Fermi level considerably well. But for the $sp^{3}$ hybridization in phosphorus one should consider at least 4 atomic orbitals for an appropriate tight-binding model. 
From an experimental point of view the multilayer structures are more convenient in laboratory because of difficulty to obtain monolayer. In the bilayer and multilayer 2D nano-structures, number of layers and stacking may tune different physical properties. The electronic band gap is tunable by stacking in silicene \cite{fu2014stacking,padilha2015free}. The optical properties is also stacking dependent in graphene \cite{wang2010stacking,koshino2013stacking} and black phosphorus \cite{shu2016stacking,ccakir2015significant}. Due to the buckling atomic structure, the blue phosphorus has more possible stacking than the graphene. The bonding between layers due to the van der Waals interaction should be considered in the DFT-D model \cite{grimme2006semiempirical}. In the tight-binding calculations, the binding between layers is modeled by additional hopping between atoms.
Here, we study the electronic and optical properties of monolayer and bilayer blue phosphorus. In the case of bilayer blue phosphorus we consider four different stacking of adjacent layers. For the electronic calculations from DFT and four atomic orbital tight-binding models were employed. Finally, the optical properties of monolayer and most stable bilayer structure are calculated based on DFT. 
\section{Model and method}
We investigate the electronic properties of monolayer and bilayer blue phosphorus with different stacking in the DFT and tight-binding theories. We fit the DFT results with tight-binding model to obtain the required parameters which are applicable for future theoretical study of bilayer blue phosphorus.  
\subsection{Density functional theory}
In this work, all the first-principles calculations are  performed  by  using
VASP package \cite{PhysRevB.54.11169}. The exchange correlation potential is approximated by generalized gradient approximation (GGA) with PBE \cite{perdew1996generalized,PhysRevLett.78.1396}. A plane-wave basis set with kinetic energy cutoff of 500 eV is used. All atomic positions and
lattice constants are optimized by using the conjugate gradient method with DFT-vdW \cite{PhysRevB.81.115126}. Moreover, Brillouin zone sampling with Monkhorst-Pack method \cite{PhysRevB.13.5188} of $ 24\times24\times1 $ $ \mathrm{k} $-points and to eliminate the interaction between monolayers in supercell, $\sim 30 $ 
\AA{} vacuum were considered. The convergence for energy was set as 10$ ^{-8} $ eV between two steps and the maximum Hellmann-Feynman forces acting on each atom was less than $ 0.001 $ eV/\AA{} upon ionic relaxation.

\begin{figure}[t]
	\includegraphics[width=0.45\textwidth,clip]{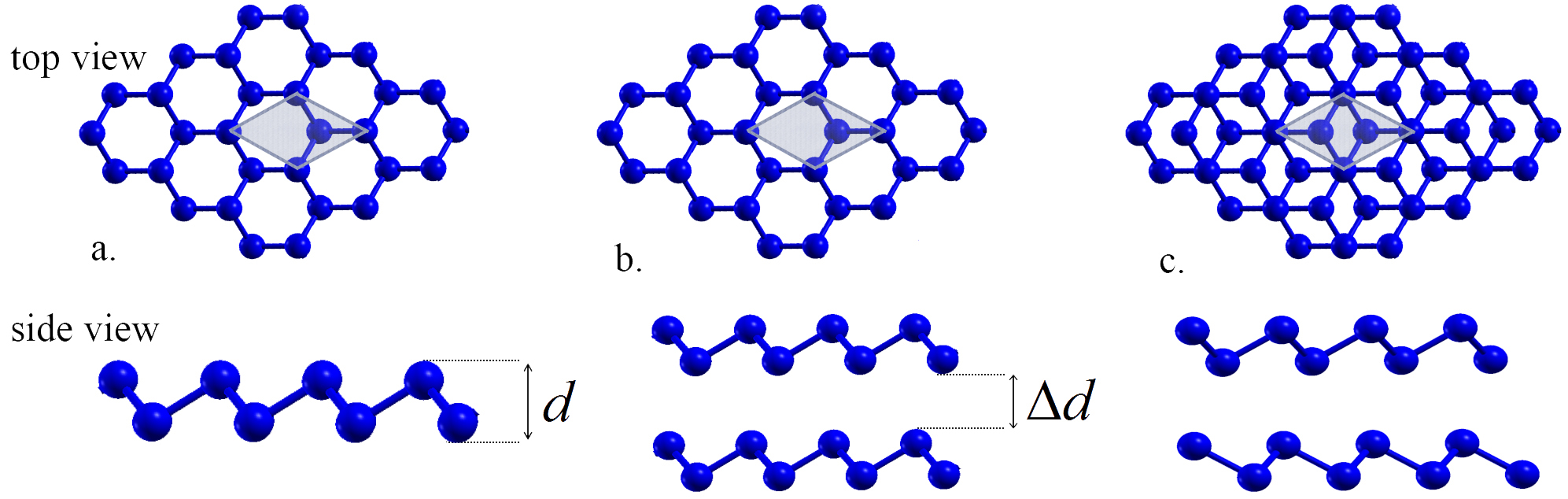}
	\caption{Structure of (a) monolayer, (b) AA stack bilayer and (c) AB stack bilayer blue phosphorus.}\label{FIG1}
\end{figure}

\subsection{Tight-binding calculations}
In the tight-binding calculations we consider four atomic orbitals per phosphorus atom as the basis set. For monolayer the hopping between nearest-neighbour (NN) and next-nearest-neighbour (NNN) are included in the tight-binding Hamiltonian. In the case of bilayer the hopping between adjacent layers is also included in the Hamiltonian. To construct the total Hamiltonian the required Slater-Koster hopping parameters which include the on-site energy of s and p atomic orbitals, hopping parameter between nearest-neighbor $t_{NN}$ and next-nearest-neighbour $t_{NNN}$ atomic sites and layers are extracted by fitting the band structure with DFT results in the first Brillouin zone. The real space Hamiltonian matrix is Fourier transformed and diagonalized to find the electronic bands as a function of wave vector in the first Brillouin zone. Here, all tight-binding calculations are performed by using a self-developed code.

\subsection{Optical properties}
To learn more about the technological importance of these structures, we focus our attention on optical properties using GGA-PBE functional. The linear response of a system due to an external electromagnetic radiation is described by the complex dielectric function $\varepsilon (\omega)$=$\varepsilon_{1} (\omega) +i\varepsilon_{2} (\omega)$ \cite{sun2004optical}. The dispersion of the imaginary part of complex dielectric function $ \varepsilon_{2} (\omega) $ was obtained from the momentum matrix elements between the occupied and unoccupied wave functions as follows,

\begin{eqnarray}
\varepsilon_{2}^{(\alpha\beta)}&=&\dfrac{4\pi^{2}e^{2}}{\Omega}\lim_{q\to 0} \dfrac{1}{q^{2}}\sum\limits_{c,v,\boldsymbol{k}}2\omega_{\boldsymbol{k}}\delta\left( \epsilon_{c\boldsymbol{k}}-\epsilon_{v\boldsymbol{k}}-\omega\right)\notag \\
&\times&\left\langle u_{c+\boldsymbol{k}+\boldsymbol{e}_{\alpha q}} \lvert u_{v\boldsymbol{k}} \right\rangle \left\langle u_{c+\boldsymbol{k}+\boldsymbol{e}_{\alpha q}} \lvert u_{v\boldsymbol{k}} \right\rangle^{*} \label{1}  
\end{eqnarray}

where the $ c$ and $ v$ correspond to conduction and valence band states respectively, and $ u_{c{\mathbf{k}}}$ is the cell periodic part of the orbitals at the k-point $ \bf k$. The real component of the dielectric function, $\varepsilon_{1} (\omega)$ is calculated via the Kramers–Kronig transformation \cite{hu2007first}. Then, other important optical constants such as the reflectivity $R (\omega) $, the electron energy-loss spectrum $L (\omega) $, as well as the refractive index $n (\omega) $, and the extinction coefficient $ k (\omega) $ were calculated using the following expressions \cite{egerton2011electron,PhysRevB.59.1776}:

 \begin{eqnarray}
R(\omega)&=&\left|\dfrac{\sqrt{\varepsilon(\omega)}-1}{\sqrt{\varepsilon(\omega)}+1}\right|^{2}, \quad  L(\omega)=\left(\dfrac{\varepsilon_{2}(\omega)}{\varepsilon_{1}^{2}(\omega)+\varepsilon_{2}^{2}(\omega)}\right) \notag \\
 n(\omega)&=&\left(\frac{\sqrt{\varepsilon_{1}^{2}(\omega)+\varepsilon_{2}^{2}(\omega)}+\varepsilon_{1}(\omega)}{2}\right)^{1/2},\notag \\ 
 k(\omega)&=&\left(\frac{\sqrt{\varepsilon_{1}^{2}(\omega)+\varepsilon_{2}^{2}(\omega)}-\varepsilon_{1}(\omega)}{2}\right)^{1/2}.\label{2}
 \end{eqnarray}%

\section{Results and discussions}
\begin{figure}[t]
\includegraphics[width=0.48\textwidth,clip]{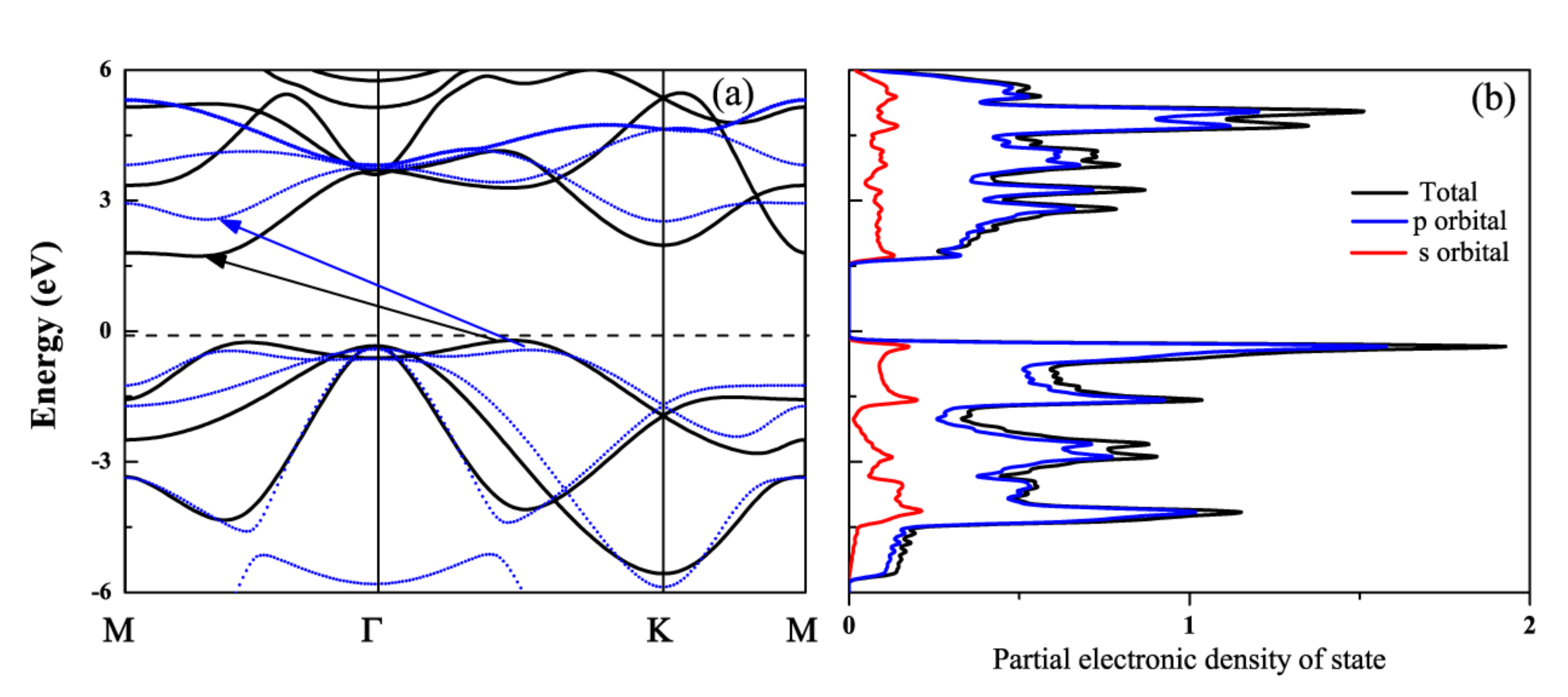}
\caption{(a)Electronic energy levels and (b) PDOS for monolayer blue phosphorus in the DFT (black solid line) and tight-binding (blue dot) theories.} \label{FIG2}
\end{figure}

The monolayer blue phosphorus consists two different sub-lattices that are separated by the buckling length as shown in FIG.\ref{FIG1}. The buckling length for monolayer is 1.23 \AA{} which is comparable with previous reports \cite{ghosh2015electric} and stanene buckling length \cite{modarresi2015effect}. All the structural parameters have also shown in Table \ref{TAB1}. The electronic band structure in the DFT and tight-binding models and partial density of states (PDOS) are plotted in FIG.\ref{FIG2} for  monolayer. The monolayer blue phosphorus is a semiconductor with indirect gap. The valence band maximum (VBM) and conduction band minimum (CBM) are between $\Gamma$-K and $\Gamma$-M in the first Brillouin zone, respectively. The gap value is 1.94 eV for DFT which is in fair agreement with tight-binding band structure. Black/blue arrows show the position of VBM and CBM in DFT/tight-binding theories. According to the PDOS for different atomic orbitals, the main contribution around the Fermi level is related to the $p$ atomic orbitals. Unlike the graphene, contribution of $s$ atomic orbitals in the total density of states is not negligible which shows the importance of different hybridization of $s$ and $p$ atomic orbitals for tight-binding calculations.   
   
\begin{figure}[t]
\includegraphics[width=0.48\textwidth,clip]{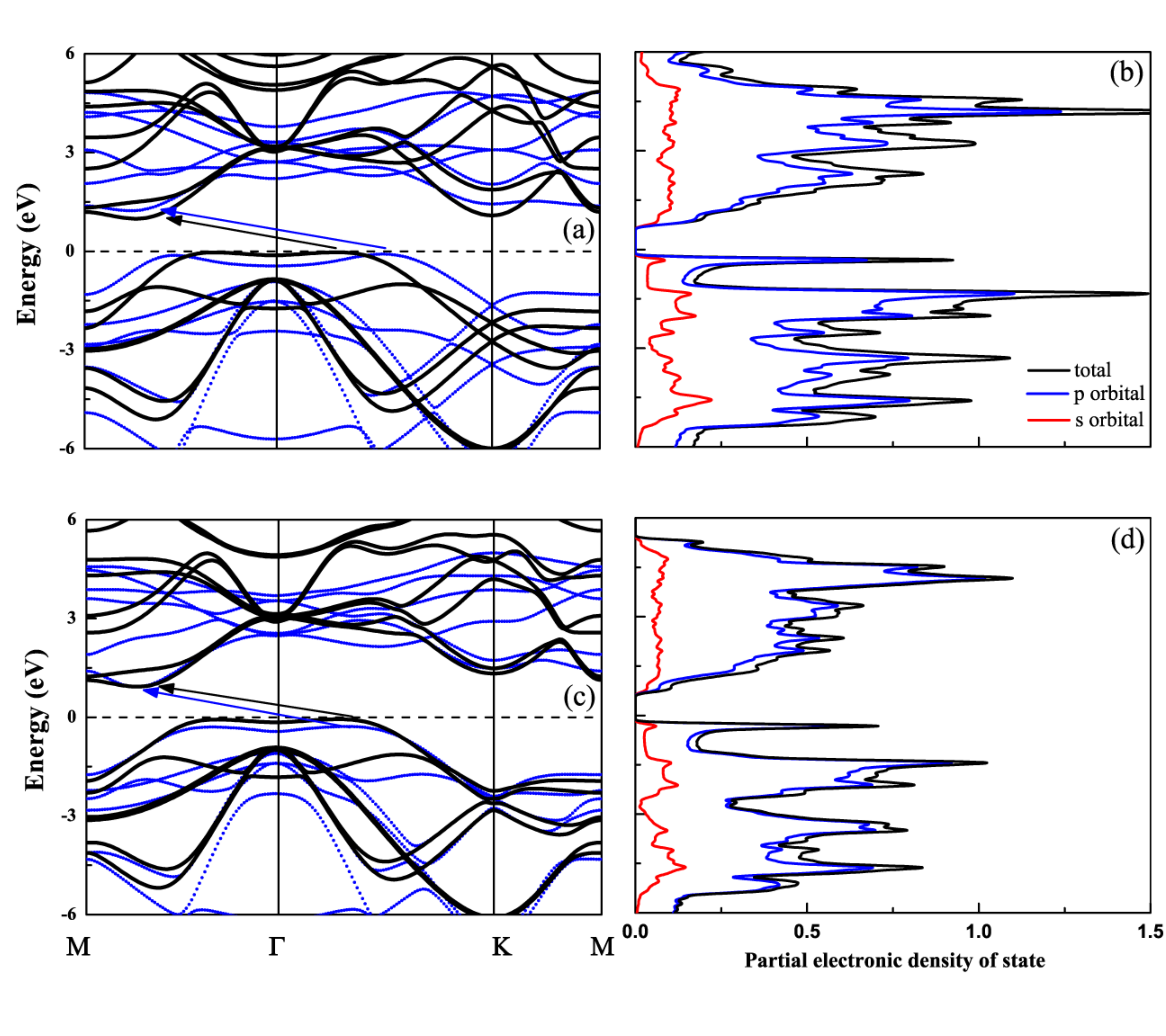}
\caption{Electronic band structure and PDOS for (a,b) AA and (c,d) AB stacks of bilayer blue phosphorus.}\label{FIG3}
\end{figure}

For bilayer blue phosphorus we examined different stacks as shown in FIG.\ref{FIG1}. In the AA stack (FIG.\ref{FIG1}(b)) the second layer is exactly above the first one but in the AB stack (FIG.\ref{FIG1}(c)) the upper layer is moved in $xy$ plane with respect to the first layer. Due to the buckling, there are different atomic configurations for AB bilayer structure. The relaxation process for AB stack is started from different configurations to guarantee the global minimum energy for AB structure. We minimized the total internal atomic force and stress for all structures. According to our calculations of the AA and AB stack bilayer blue phosphorus have the minimum energy and considered as the most stable structures in the following. The inter-layer binding energy for AA and AB stacks is 25 meV (cohesive energy as 12.5 meV/atom) which is comparable for a typical van der Waals layered structure, such that cohesive energy of graphene-hexagonal boron nitride superlattices was found around $\sim 9.5$ meV/atom from GGA+vdW functional \cite{kaloni2012electronic}. The weak binding between layers in bilayer blue phosphorus make it possible to exfoliate 2D layer from the bulk one. The inter-layer distance in bilayer structures are 3.24 and 3.21 \AA{} for AA and AB stacks, respectively. Buckling parameter of bilayer is almost same as monolayer for blue phosphorus. Electronic band structure and PDOS of bilayer blue phosphorus plotted in FIG.\ref{FIG3} for AA and the stable AB stacks.

The position of VBM and CBM are hardly ever changed in the bilayer structure with respect to the monolayer but they move toward each other that decreases energy gap to $ \sim $1 eV. In the bilayer structures the atomic $p$ orbital enter to the electronic gap region. Each energy band in the monolayer is split to two bands due the interaction between two adjacent layer. The fitting process between DFT and tight-binding is done in an iterative Monte Carlo method to obtain the best possible set of parameters. The difference between on-site energy of $s$ and $p$ atomic orbitals is -4.55 eV for monolayer and different type of bilayer structures of blue phosphorus. Table \ref{TAB2} contains all the tight-binding parameters required to construct the Hamiltonian for different structures. Our tight-binding parameters for monolayer blue phosphorous are in good agreement with Ref.\onlinecite{lee2015two}. Due to the simple atomic structure for the AA stack only nearest neighbor hopping leads to relatively good results but for AB configuration we consider both nearest neighbour and next nearest neighbour hopping integrals for bilayer blue phosphorous. 

\begin{table}[h!]
	\centering
	\begin{tabular}{||c c c c c||} 
		\hline
		structures & length (\AA{}) & angle (Degree) & $ d $(\AA{}) & $\Delta$$ d $ (\AA{})\\ [0.5ex] 
		\hline\hline
		monolayer & 2.26 & 93.07 & 1.23 &  \\
		AA stack bilayer & 2.26 & 93.11 & 1.23 & 3.24\\
		AB stack bilayer& 2.26 & 93.21 & 1.23 & 3.21\\
		 [1ex] 
		\hline
	\end{tabular}
	\caption{Equilibrium structure parameters of the monolayer and bilayer blue phosphorus.}
	\label{TAB1}
\end{table}
 
\begin{table}[h!]
\centering
\begin{tabular}{||c c c c c||} 
 \hline
 Parameter & t$_{ss\sigma} $ & t$_{sp\sigma} $ & t$_{pp\sigma} $ & t$_ {pp\pi} $\\ [0.5ex] 
 \hline\hline
 monolayer$_{NN}$ & -1.0 & -2.9 & 3.3 & -0.7 \\
 monolayer$_{NNN}$ & 0.25 & -0.3 & 1.15 & -0.4 \\
 AB$_{NN}$ & -0.06 & 0.06 & 1.35 & -0.45\\
 AB$_{NNN}$ & -0.04 & 0.08 & -0.66 & -0.35\\
 AA$_{NN}$ & -0.22 & -0.35  & -1.51  & -0.3 \\ [1ex] 
 \hline
\end{tabular}
\caption{Tight-binding parameters of blue phosphorus for monolayer and bilayer in AA and AB stacks.}
\label{TAB2}
\end{table}

The PDOS for AA and AB stacks shows the contribution of s atomic orbitals in the total density of states in bilayer structure. 

\begin{figure}[t]
\includegraphics[width=0.5\textwidth,clip]{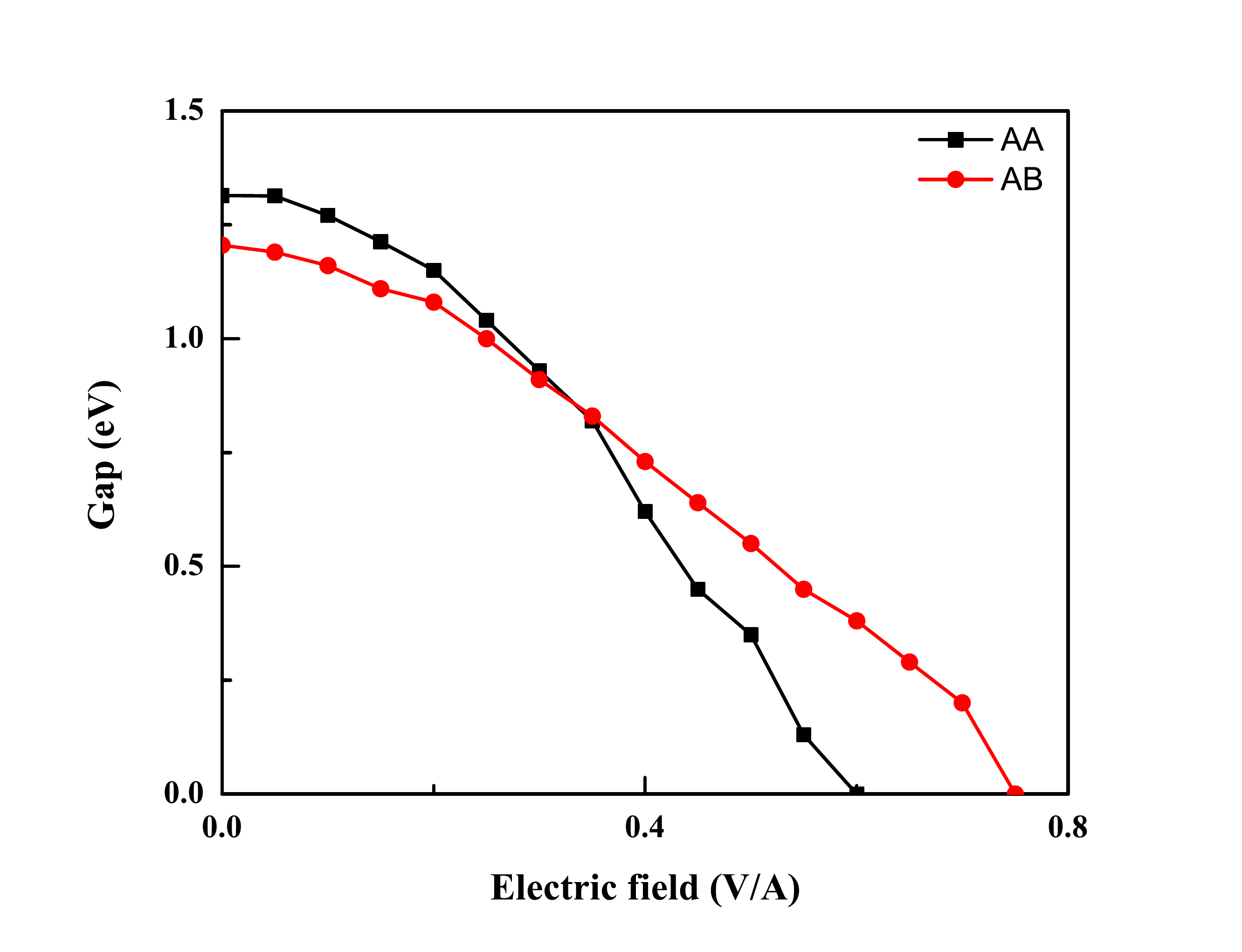}
\caption{Electronic band gap for AA and AB stacks as a function of applied external electric field.}\label{FIG4}
\end{figure}

\begin{figure}[t]
	\includegraphics[width=0.50\textwidth,clip]{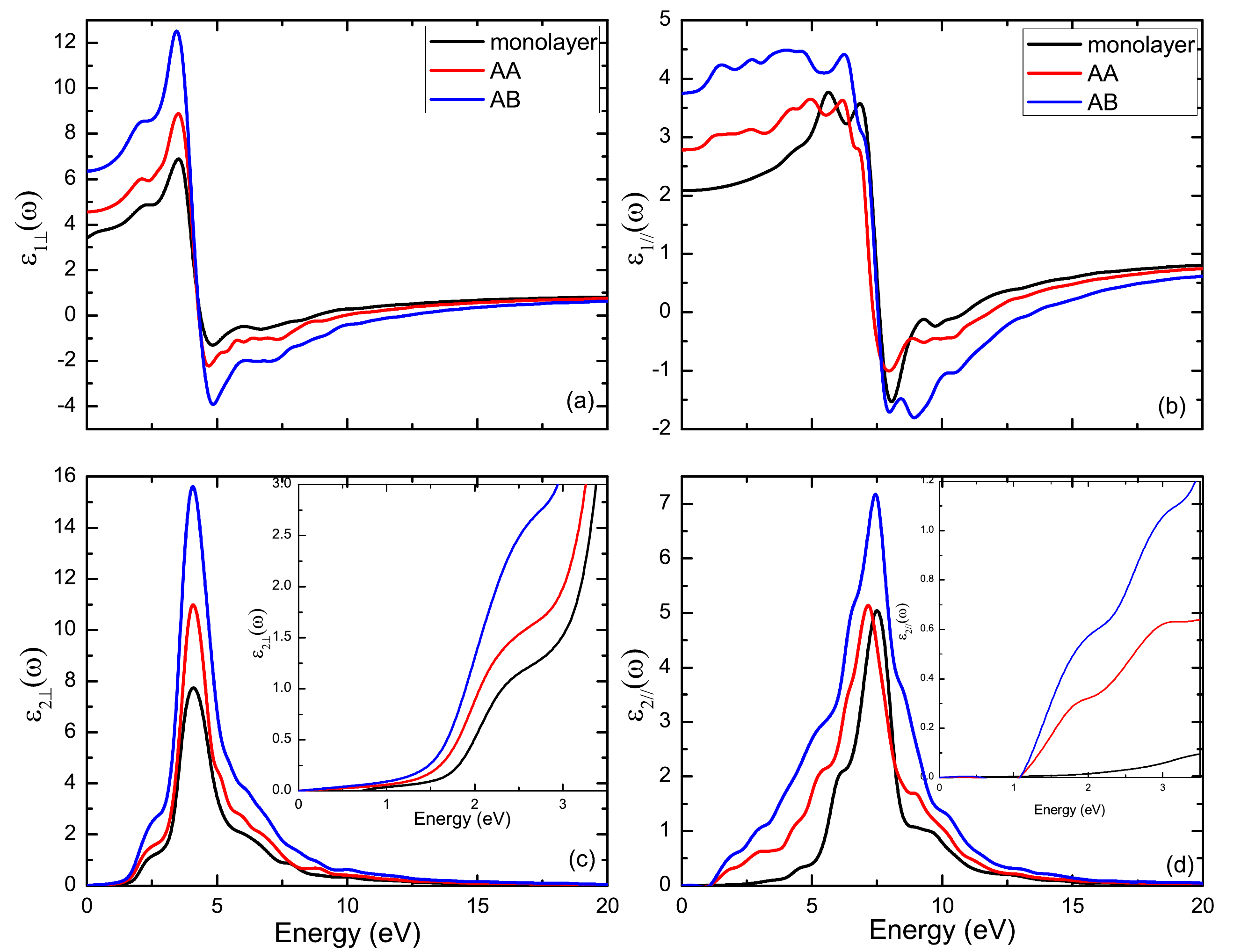}
	\caption{The computed (a),(b) real and (c),(d) imaginary part of the dielectric function of the structures versus photon energy.}\label{FIG5}
\end{figure}

The tight-binding model predicts position and size of electronic band gap that is a remarkable success for a simple atomic-orbital basis model. We examined the effect of perpendicular external electric field on the band gap of bilayer blue phosphorus in the tight-binding model. The electric field produces a potential difference on each plane and shifts energy bands related to different layers. This shift fills energy region between VBM and CBM and closes the electronic band gap for enough high electric potential. It was shown the external electric field may open band gap in other 2D material \cite{kaloni2014large}. Also the adsorption of molecules on silicene can be thought as an internal electric field that modify the band gap of structure \cite{kaloni2013quasi}. FIG.\ref{FIG4} shows the variation of band gap as a function of applied electric field for AA and AB stacks. The external electric field does not change the position of VBM and CBM for both structures but decreases the electronic gap. The energy gap is closed for electric field around E=0.6 V/\AA{} which is compatible with recent published DFT+HSE06 results \cite{ghosh2015electric}.
We now focus our attention on discussion of optical properties of monolayer and bilayer blue phosphorus. The dielectric constant is a complex function of incident photon energy. The calculated $ \varepsilon_{1} (\omega) $ and $ \varepsilon_{2} (\omega) $ parts of the electronic dielectric function for the monolayer and bilayer blue phosphorus in the range of 0-20 eV are shown in FIG.\ref{FIG5}. 

The crystal structure of blue phosphorus is hexagonal and characterized by two independent tensor components (perpendicular and parallel to z-axis) of the dielectric tensor. The static perpendicular real part of the dielectric function, $ \varepsilon_{1\perp}(0) $ are found to be 3.411, 4.551 (6.347) for monolayer and AA (AB) stack bilayer blue phosphorus, respectively. On the other hand, tha static parallel real part of dielectric function $ \varepsilon_{1\parallel}(0) $, 2.081, 2.776 (3.746) for monolayer and AA (AB) stack bilayer blue phosphorus, respectively. One can notice that from monolayer to bilayer the peaks in the $ \varepsilon_{1}(\omega) $ increase and shift to the low energy region. Due to the absence of absorption in the energy gap region, the imaginary part of dielectric function which is proportional to absorption spectra is zero in low photon energy region. The imaginary part of dielectric function depends on the polarization of incident light. For polarization perpendicular to the phosphorus plane, monolayer and bilayer  structures are almost transparent to light between 0-2 eV as shown in the inset of FIG.\ref{FIG5}c. 

For the parallel polarized light, the imaginary dielectric function of AA and AB stacks have a red shift and is more intense with respect to monolayer blue phosphorus which is related to interlayer interaction in bilayer structure. The difference between absorption of parallel polarized light may be used in laboratory to distinguish between monolayer and bilayer blue phosphorus.
Using $ \delta\varepsilon=\left(\varepsilon_{1\parallel}(0)- \varepsilon_{1\perp}(0)/ \varepsilon_{total}(0)\right) $ relation \cite{reshak2014electronic}, we calculated uniaxial anisotropy and  found to be about -0.242 and -0.242 (-0.258) for monolayer and AA (AB) stack bilayer blue phosphorus, respectively. It means that difference between perpendicular and parallel real part of dielectric function suggests anisotropic behavior of optical property. The anisotropy of optical absorption in blue phosphorus is originated from 2D nature of atomic configuration and decreased in bilayers for low energy photons. 

For larger frequencies than about 4.34 (4.32 for AA and 4.34 for AB) eV and 7.65 (7.3 for AA and 7.5 for AB) eV for the monolayer blue phosphorus, the real part becomes negative for $ \varepsilon_{1\perp} $ and $ \varepsilon_{1\parallel} $, respectively. As it can be seen from FIG.\ref{FIG5}(c), and (d), these structures have one major peaks. The highest peak of the imaginary part of the dielectric function $ \varepsilon_{2\perp} $ is located at 4.05 (4.07 for AA and 4.05 for AB) eV and $ \varepsilon_{1\parallel} $ is also found to be 7.48 (7.17 and 7.43 for AB) eV for monolayer blue phosphorus, respectively, which are related to inter-band transitions between the valence and conduction bands. In comparison the peak of bilayer blue phosphorus for both AA and AB stack are higher than monolayer blue phosphorus peak. As it is seen in FIG.\ref{FIG5}, the imaginary part of the dielectric functions for monolayer and bilayer blue phophorus within the energy range of 0-20 eV are clearly related to the their band structures that indicates the absorption behavior so that the electronic transitions from valance to conduction bands have contribution to the main part of the optical spectra. Considering the imaginary part of the parallel dielectric function, $ \varepsilon_{2\parallel}$, one can observe that the threshold energies of the dielectric function is around $\sim 1.9$ eV for monolayer, and $\sim 1$ eV for both AA and AB stack blue phosphorus. The threshold energies of the parallel dielectric function correspond to the band gaps of the systems. The threshold energy of transition between the highest valance band and the lowest conduction band is known as the fundamental absorption edge. The other peaks are related to different electronic transitions from occupied states (valance bands) to the unoccupied states (conduction bands). It should be considered that these peaks are not only been occurred from the electronic transitions between the two bands but also from a combination of direct and indirect inter-band transitions. In addition, the low energy peaks are caused by the near-band transitions.

\begin{figure}[t]
	\includegraphics[width=0.50\textwidth,clip]{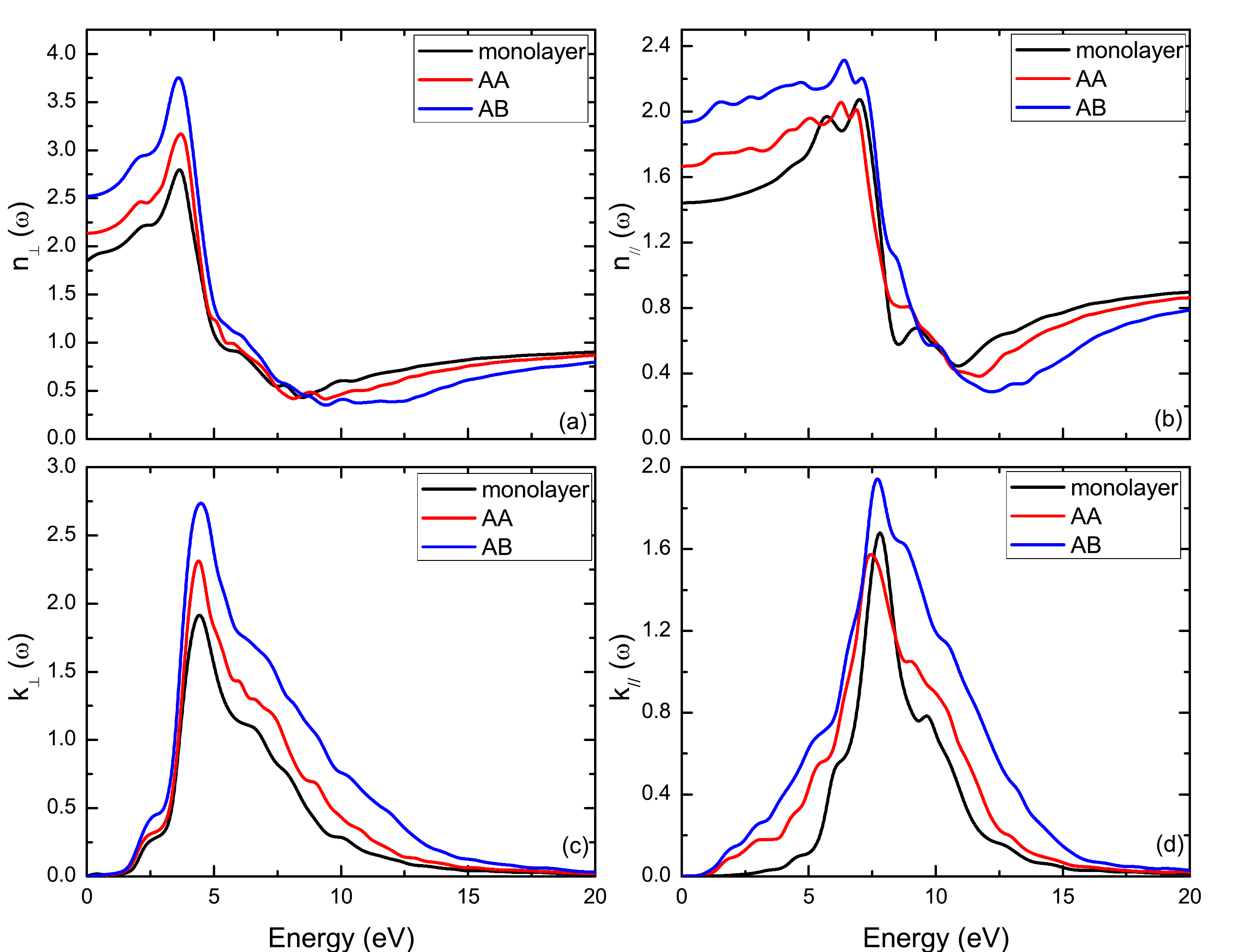}
	\caption{The computed (a),(b) refractive index $ n(\omega) $ and (c),(d) extinction coefficient $ k(\omega) $ of the structures versus photon energy.}\label{FIG6}
\end{figure}

\begin{figure}[t]
 	\includegraphics[width=0.50\textwidth,clip]{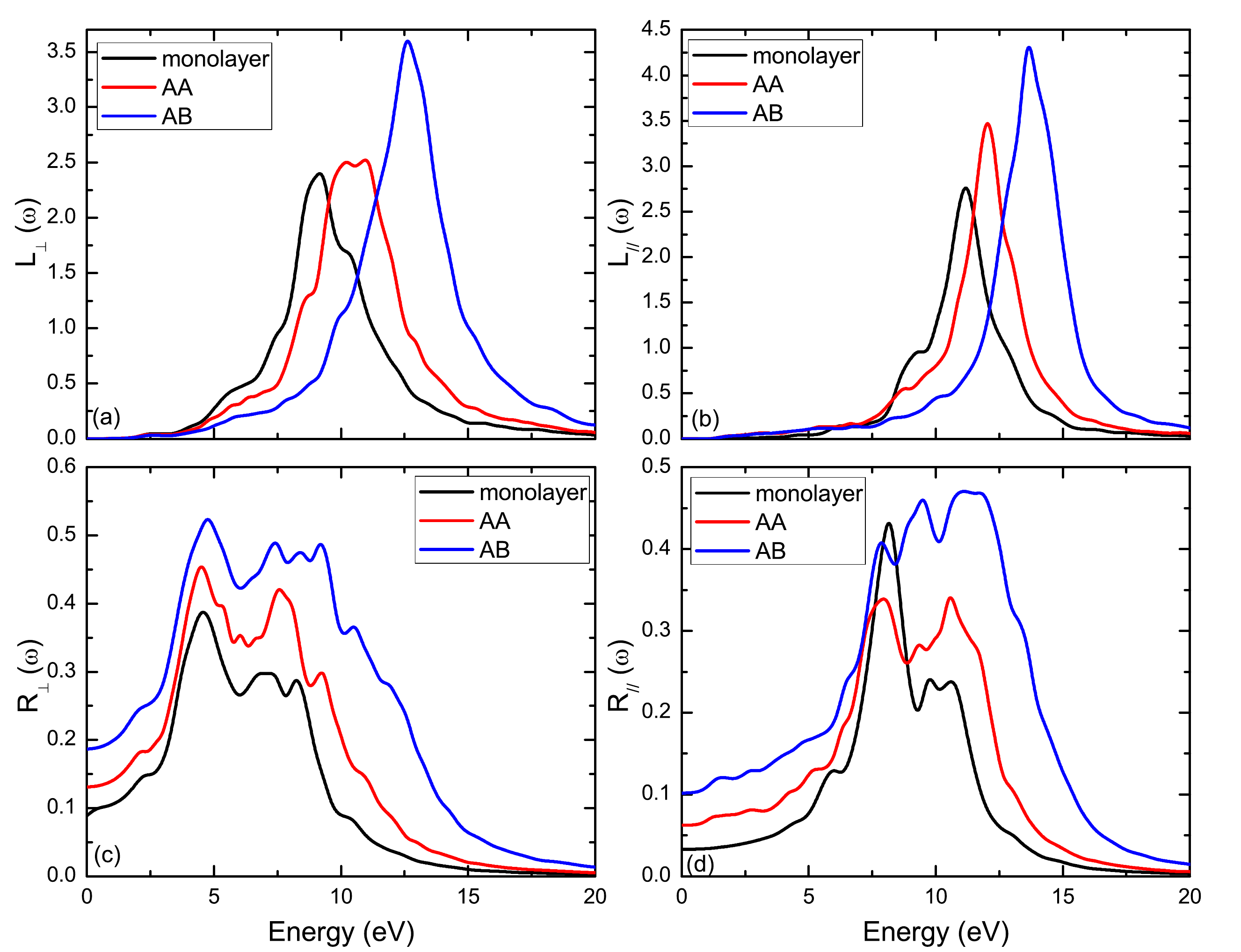}
 	\caption{The computed (a),(b) electron energy loss spectrum $ L(\omega) $ and (c),(d)  reflectivity $ R(\omega) $ of the structures versus photon energy.}\label{FIG7}
 \end{figure}

The calculated refractive index $ n(\omega) $, extinction coefficients $ k(\omega) $, energy loss function $ L(\omega) $ and reflectivity $ R(\omega) $ are estimated by Kramers-Kronig relations[3] and given in Eq.(\ref{2}). Our obtained results are plotted in FIGs. \ref{FIG6} and \ref{FIG7}. The calculated refractive index is displayed in FIG.\ref{FIG6}(a) and (b) for monolayer, AA and AB stack blue phosphorus. While the predicted values of perpendicular static refractive index $ n_{\perp}(0) $ are 1.84, 2.13, and 2.52, the parallel static refractive index values $ n_{\parallel}(0) $ are 1.44, 1.66 and 1.93 for  monolayer, AA and AB stack bilayer blue phosphorus, respectively. The static parallel refractive index $ n_{\parallel}(0) $=1.44 for monolayer blue phosphorus comparable with graphene  ($n_{\parallel}(0) $=1.12 and $ n_{\perp}(0) $=2.75) \cite{Rani201428} and 2D-ZnS ($n_{\parallel}(0) $=1.66) \cite{Lashgari201676}. The main peak values of refractive index for monolayer, AA and AB stack bilayer blue phosphorus are 2.79 at 3.60 eV, 3.17 at 3.70 eV, and 3.75 at 3.60 eV,  respectively. From the FIG.\ref{FIG6} (c) and (d), we have predicted the extinction coefficients for monolayer and AA (AB) stack bilayer blue phosphorus to be 1.902 and 2.30 (2.74), respectively. The extinction coefficients are needed to calculate for absorption and corresponds also to transmission of light that allows the experiments by using optical spectrometers. As shown in FIG.\ref{FIG6} (c) and (d), maximum values of extinction coefficients in perpendicular direction are AB stack (2.74) $>$ AA stack (2.30) $>$ monolayer (1.9) blue phopshorus while in parallel direction Ab stack (1.9) $>$ monolayer (1.7) $>$ AA stack (1.6). It means that the threshold energies would have been different and dependent to the parallel or perpendicular directions. 

The black phosphorus solves the high dark current problem in graphene photodetectors \cite{youngblood2015waveguide} and used in field-effect transistors \cite{buscema2014fast}. Due to the electronic and optical band gap the blue phosphorus may be the next candidate for application in optoelectronic devices.  The absorption coefficients is calculated by extinction coefficient, $\alpha (\omega)=4\pi $k$(\omega)/\lambda$, where $\lambda$ is the photon wavelength. According to our calculations the absorption coefficients for visible light region is in order of $\sim 10^{5}$ cm$^{-1}$ which is comparable by silicon absorption \cite{sze2006physics,zacharias2015stochastic}. Also the difference between phonon spectra gap and the hardest acoustic mode is much bigger in the blue phosphorus with respect to black phosphorous \cite{zhu2014semiconducting} which prevents Klemens decay \cite{yu2015topological} for high efficient 2D solar cell applications.

One further point of interest is energy loss functions, which is an important factor describing the energy loss of a fast electron traversing in a material, as depicted in FIG.\ref{FIG7} (a) and (b) for monolayer and bilayer blue phosphorus. The electrons of solids could be excited in several ways. One of them has been done as, when a fast electron passes through a solid, it may has been loss some energy, known as $L(\omega)$, and excites the electrons of the solid. Inter and intra-band transmissions, plasmon excitations along with other possible ones contribute to forming energy loss spectrum, therefore all excitations could be identified by analyzing energy loss spectrum which is related to dielectric function and given in Eq.(\ref{2}). Energy loss spectrum peaks are related to not only inter-band transitions but also corresponded to the plasmons that are collective oscillations of free electrons with energies dependent to the density of valance electrons. The maximum peaks in the energy-loss function indicate that plasmon resonance occurs at around 11.205 and 12.054 (13.634) eV for monolayer and AA (AB) stack bilayer blue phosphorus, respectively. It can be pointed out that the plasma frequency of AB stack bilayer blue phosphorus is the largest one. Reflectivity $R(\omega)$, is an important quantity to determine the optical properties which is mentioned in Eq.(\ref{2}). FIG.\ref{FIG7} illustrates the reflectivity spectrum for monolayer and bilayer systems of blue phosphorus. The $R(\omega)$ curve for all structures have a main peak and the reflectivity tends to zero for high energy photons. The peaks have been occurred from the inter-band transitions. The static parallel reflectivity $ R_{\parallel}(0) $ values are higher than the static perpendicular reflectivity $ R_{\perp}(0) $ values for bilayer systems while the static perpendicular reflectivity $ R_{\perp}(0) $ value is higher than $ R_{\parallel}(0) $ value for monolayer system. As it can be seen from FIG.\ref{FIG7}(c) and (d), while the static perpendicular reflectivity $ R_{\perp}(0) $ is 0.088 and 0.131 (0.186), the static parallel reflectivity $ R_{\parallel}(0) $ are 0.0328 and 0.0625 (0.101), and the maximum values of that are about 0.38 at 4.647 eV and 0.45 at 4.506 eV (0.426 at 8.171 eV) for monolayer and AA (AB) stack bilayer blue phosphorus, respectively. Among these structures, AB stack bilayer blue phosphorus shows the highest reflectivity at low energy due to its more pronounced metallicity character \cite{singh2013first}.

\section{Conclusion}
In summary, we study the electronic and optical properties of monolayer and bilayer blue phosphorus in AA and AB stacks. The comparison between DFT and Slater-Koster tight-binding provides table of hopping parameters for each atomic configuration. The weak binding between layers proposed the possibility of exfoliation 2D blue phosphorus from bulk in laboratory. Based on the tight-binding model an external perpendicular electric field produces atomic dependent potential that closes electronic band gap in bilayer blue phosphorus. Finally, we reported the stacking dependent optical properties in bilayer blue phosphorus by using DFT. To compare, the static parallel refractive index $ n_{\parallel}(0) $=1.44 and the static perpendicular refractive index $ n_{\perp}(0) $=1.84 for monolayer blue phosphorus comparable with graphene ($n_{\parallel}(0) $=1.12 and $ n_{\perp}(0) $=2.75). In perpendicular direction, the refractive index value for blue phosphorene is less than graphene as expected. The blue phosphorus may has the potential application in future (opto)electronic devices based on 2D materials. 

\section*{References}
\bibliography{referans}
\bibliographystyle{apsrev4-1}

\end{document}